# A passively tunable non-resonant acoustic metamaterial lens for selective ultrasonic excitation

H. Zhu, F. Semperlotti

*Department of Aerospace and Mechanical Engineering, University of Notre Dame, Notre Dame, Indiana, 46556*

**Abstract**

In this paper, we present an approach to ultrasonic beam-forming and beam-steering in structures based on the concept of embedded non-resonant acoustic metamaterial lenses. The lens design exploits the principle of acoustic drop-channel which enables the dynamic coupling of multiple ultrasonic waveguides at selected frequencies. In contrast with currently available technology, the embedded lens allows generating directional excitation by means of a single ultrasonic transducer. The lens design and performance are numerically investigated by using Plane Wave Expansion and Finite Difference Time Domain techniques applied to bulk structures. Then, the design is experimentally validated on a thin aluminum plate where the lens is implemented by through-holes. The dynamic response of the embedded lens is estimated by reconstructing, via Laser Vibrometry, the velocity field induced by a piezoelectric shaker source.

The development of highly focused and directional excitation is an area of major interest in the Structural Health Monitoring (SHM) field. The ability to send ultrasonic energy in a preferential direction results in improved damage sensitivity and resolution due to the increased interaction between the damage and the interrogation signal. Highly directional excitation is also of importance in damage detection of highly directional or anisotropic material, such as layered composites [1], where the direction of energy propagation can be largely different from the initial direction of the emitted wave. Under these conditions, directional interrogation can help compensating for this intrinsic characteristic of the material. Directional interrogation is also an enabling technology to address multiple damage scenarios. The ability to selective scan a structural element and acquire data from individual damage can drastically improve the accuracy of the detection. To-date, Phased Array (PA) technology [2,3] has been among the most successful techniques to achieve ultrasonic beam forming for SHM applications. PA exploits a set of transducers network activated following pre-defined time delays in order to either achieve a directional wave front or focused excitation at a prescribed spatial location. Despite being a simple and robust method, PA has two main limitations which prevent its extensive use in practical applications. First, PA technology requires a large number of transducers in order to produce focused and steerable excitation. From the SHM perspective, this aspect is considered a major downside because it results in increased probability of malfunction and false alarms as well as in higher system complexity that ultimately affects fabrication and installation (e.g. harnessing, power source, etc.). The second drawback of PA is related to its inability to generate collimated signals since either directional or focused excitation is the result of constructive/destructive interference produced by the linear superposition of multiple omni-directional wavefronts. From a SHM perspective, non-collimated waves are less suitable due to the increase in the background noise resulting from the boundary and damage induced back-scattering from the side lobes of the excitation signal. A viable alternative to PAs was recently proposed by the authors [4]. The concept is based on the use of a resonant acoustic metamaterial lens surrounding a single piezoelectric transducer. The lens is fully embedded in the structure so that the structure itself becomes an integral part of the transducer system. The lens is designed in order to generate either focused or collimated ultrasonic beams that can be steered in different directions by simply tuning the frequency of excitation. The design exploits resonant inclusions arranged in an anisotropic lattice structure which results in hyperbolic equi-frequency-contours that are at the basis of the focusing and collimation mechanism. Despite offering very interesting dynamic performance, the resonant lens has higher design and fabrication complexity which make this design suitable only for limited sections of the lens structure.

In this letter, we present an alternative design to embedded acoustic lenses based on non-resonant metamaterials [5,6]. Compared to the resonant lens, the new design delivers a much lower fabrication complexity while still achieving directional and steerable excitation with a single ultrasonic transducer. Note that, despite a lower fabrication complexity, the non-resonant design cannot provide the entire spectrum of dynamic performance of its resonant counterpart as it will be further discussed later on.

The proposed design is based on the concept of dynamic structural tailoring of the host structure achieved via drop-channel design. The drop-channel [7-10] is a simple and useful concept in metamaterial design that allows tailoring the wave propagation characteristics of the host material by exploiting a carefully engineered network of defects. Both point and line defects (also called, cavity and waveguide defects) can be properly sized and distributed across the material in order to tailor the dispersion characteristics in both the frequency and spatial domains. This property can be exploited to create a variety of phenomena including preferential directions of energy propagation activated at selected frequencies as well as embedded mechanical bandpass filters [11]. In the following, the operating principle at the basis of the non-resonant lens design is first illustrated numerically by showing how the dispersion characteristics of a bulk structure can be tailored to yield selected propagation performance. Then, the concept of acoustic lens is experimentally validated on an aluminum thin plate-like structure with an embedded semicircular lens.

We consider a bulk perfectly periodic elastic metamaterial with a square lattice structure (lattice constant $a$). The material is assumed made of an epoxy background and carbon cylindrical inclusions having radius $r_s=0.418a$. The dispersion characteristics of this material are calculated using the supercell [12] Plane Wave Expansion (PWE) [13,14] method and plotted along the boundary of the first Brillouin zone (FIG. 1a). For the sake of clarity and without loss of generality, the band structure is shown only for the Shear Vertical (SV) mode which corresponds to a particle displacement parallel to the axis of the cylindrical inclusions. FIG. 1a shows the existence of a full bandgap [15] for the SV mode in the non-dimensional frequency range $\Omega = \omega a/2\pi c_t = 0.14 \div 0.2$, where $c_t$ is the transverse phase velocity of carbon. In this frequency range propagating waves are not supported, therefore the perfectly periodic material effectively acts as a mechanical stopband filter. Wave propagation at frequencies inside the bandgap can be achieved by creating a network of defects associated with either localized or partially spatially confined modes [16,17]. The term defect typically refers to the lack of periodicity of the host material that can be achieved, as an example, by altering the geometric or the material properties of one or more inclusions. The defect type and size can be used to control the dynamic response of the localized mode and the way different defects interact with each other. The most common defect types are point and line defects. From a general perspective, point defects create highly spatially localized (standing wave type) modes while line defects act as waveguides allowing propagation in the direction of the channel [18]. An example of point and line defects is provided in FIG. 1(b) and (c) showing the generation of defect modes (red lines) in the carbon-epoxy host structure. The insets also show the typical localized mode corresponding to one of these modes.

Unique wave paths can be created inside the host by properly designing a network of structural defects interconnected at selected frequencies. FIG. 2 shows a conceptual example of a defect network where two adjacent waveguides are coupled at selected frequencies (inside the bandgap) by exploiting the cavity modes of a single point defect (FIG. 2b inset). Since the frequency of the localized modes can be tuned by designing the properties of the defect, it follows that the propagation characteristics of the metamaterial can be tailored in both the spatial and frequency domain by properly designing a network of defects.

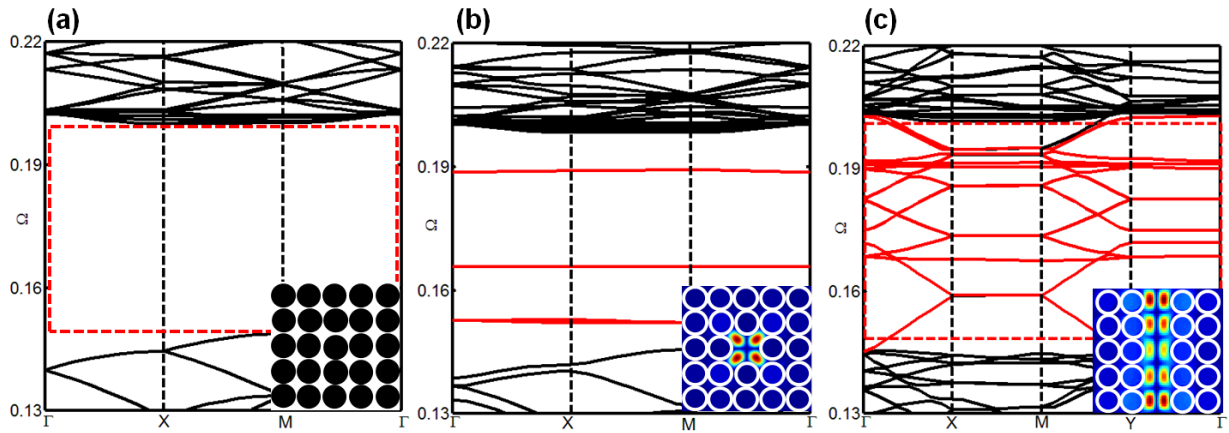

FIG. 1. Band structure of the SV mode along the first Brillouin zone for the (a) perfectly periodic, (b) point-defected, and (c) line-defected bulk crystal. Numerical results show the existence of a full bandgap ((a) red dashed box) and the generation of localized modes ((b) and (c) red solid curves) inside the bandgap, respectively. The insets in (a)-(c) show the perfect crystal, and the localized displacement fields associated with the defect modes.

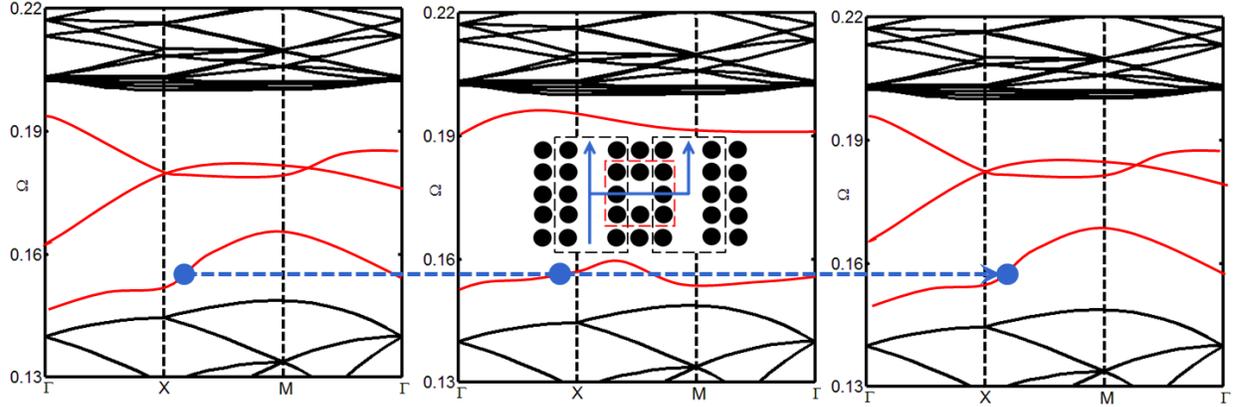

FIG. 2. Conceptual illustration of the drop-channel coupling mechanism for the design of the non-resonant acoustic lens. The schematics show the bandgap for a perfectly periodic material and the conceptual defect modes (red lines) that can be created by engineering line (left and right) and point (center) defects. When cavity and waveguide modes are supported at the same frequency the dynamic coupling between the adjacent waveguides is largely enhanced and vibrational energy is effectively exchanged from one guide to the other.

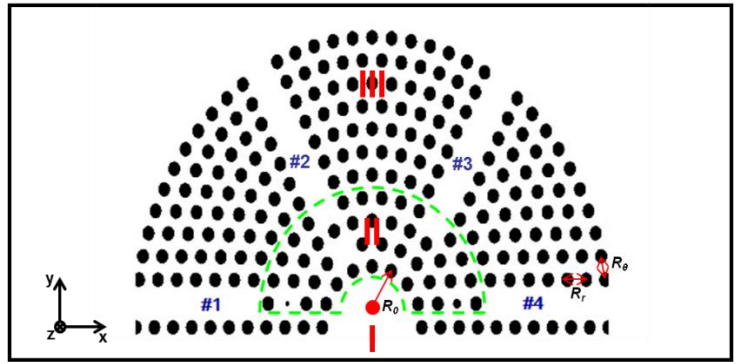

FIG. 3. Schematic of the non-resonant acoustic lens. The lens is made up of three regions: (I) an inner section represented by a semi-circular area without inclusions. This area hosts the ultrasonic transducer and delivers the excitation. (II) An intermediate section (inside the dashed box) containing the point defects to achieve the dynamic coupling with the channels. (III) An outer section that hosts the waveguides oriented in different directions. Selected drop-channels can be activated at selected frequencies inside the bandgap.

The drop-channel concept can be exploited to design acoustic lenses embedded in a supporting structure and able to achieve beam-forming and beam-steering using a single ultrasonic transducer. In order to validate the design and assess the performance of the acoustic lens, we numerically investigated the dynamic response of a bulk material with an embedded semicircular lens. The lens was assumed made of nickel cylindrical inclusions embedded in an epoxy matrix. The inclusions were distributed according to the geometry shown in FIG. 3. The lens is mostly divided in three sections: (I) an inner section without cylindrical inclusions, (II) an intermediate section with embedded point defects, and (III) an outer section containing a set of radial waveguides at prescribed azimuthal locations. The inner section is intended to host the source of the ultrasonic excitation, the waveguides are used to form the ultrasonic beam, while the intermediate section provides dynamic coupling at selected frequencies between the source and the waveguides. To test the performance of the acoustic lens, the bulk material was modeled using Navier's equations for a linear elastic space:

$$\rho(x,y,z)\ddot{u}_i(x,y,z,t) = \partial_i\left[C_{ijkl}(x,y,z)\partial_k u_l(x,y,z)\right] \quad (1)$$

where $\rho$ and $C_{ijkl}$ are the space-dependent density and elastic tensor. As previously mentioned, we consider only the SV mode therefore the governing equations for an isotropic material take the form:

$$\rho \frac{\partial^2 u_3}{\partial t^2} = \frac{\partial \tau_{xz}}{\partial x} + \frac{\partial \tau_{xz}}{\partial y} \quad (2)$$

$$\tau_{xz} = C_{44} \frac{\partial u_3}{\partial x} \quad (3)$$

$$\tau_{yz} = C_{44} \frac{\partial u_3}{\partial y} \quad (4)$$

Eqn. (2) can be solved using the Finite Difference Time Domain (FDTD) [19] method which results in the following discretized form of the governing equations:

$$u_3(i,j,t+1) = \frac{\Delta t^2}{\rho(i,j)\Delta x}[\tau_{xz}\left(i+\frac{1}{2},j,t\right) - \tau_{xz}(i-\frac{1}{2},j,t)] + \frac{\Delta t^2}{\rho(i,j)\Delta x}[\tau_{xz}\left(i+\frac{1}{2},j,t\right) - \tau_{xz}(i-\frac{1}{2},j,t)] \\ + 2u_3(i,j,t) - u_3(i,j,t-1) \quad (5)$$

$$\tau_{xz}(i+\frac{1}{2},j,t) = C_{44}(i,j)[u_3(i+1,j,t) - u_3(i,j,t)] \quad (6)$$

$$\tau_{xz}(i,j+\frac{1}{2},t) = C_{44}(i,j)[u_3(i,j+1,t) - u_3(i,j,t)] \quad (7)$$

where $\rho$ and $C_{44}$ are the density and the spatially dependent elastic constant, $u_3$ is the particle displacement at integer nodes, $\tau_{xz}$ and $\tau_{yz}$ are the two shear components of the stress tensor involved in the SV mode. Perfectly Matched Layers [20] (PML) were used to simulate an infinite domain and to reduce the effect of reflected waves. A point source excitation is located at the center of the lens to simulate the transducer. The geometric and material parameters used for the calculations are: for epoxy $\rho$ =1180 kg/m³ and $C_{44}$=1.6 GPa; for nickel $\rho$ =8905 kg/m³ and $C_{44}$ =80 GPa. The radius of the inner semicircular area is $R_0$ = 2a. The spacing between the inclusions is $R_r$ = 0.75a in the radial direction and $R_\theta = \frac{\pi}{4}a$ in the azimuthal direction in order to obtain straight waveguides. The lattice constant $a$ is set to 3 cm. The size of the inclusion forming the semi-circle geometry is set to $r$ = 0.24a. The drop-channels were created by removing arrays of inclusions at $\theta_1$=0°, $\theta_2$=60°, $\theta_3$=120°, and $\theta_4$=180°. The channels were coupled to the source by using point defects in the intermediate section having radius of 0.33r, 0r, 0.99r and 0.66r, respectively. These point defects resulted in cavity modes tuned at $f_1$=20.87 kHz, $f_2$=24.96 kHz, $f_3$=42.35 kHz, and $f_4$=41.17 kHz. Note that each point defect is associated with multiple localized modes although the coupling is designed for a specific frequency. The modes were selected according to two main criteria: 1) largest spectral peak amplitude, and 2) separation of the activation frequencies. The performance of the lens was tested by injecting a single tone harmonic excitation tuned at the frequencies of the cavity modes (i.e. $f_1$ to $f_4$) and applied at the geometric center of the lens. Numerical results (FIG.4 (b)-(d)) are presented in terms of maps of the squared normalized displacement field $\left|\frac{u_3}{\max(u_{3\_exc})}\right|^2$ (normalized w.r.t. the maximum displacement at the driving point) and directivity plots. Results confirm that the lens design based on the drop-channel concept is able to generate directional ultrasonic excitation by using a single transducer. The excitation direction can be steered by tuning the frequency of excitation to the frequency of the cavity mode corresponding to the desired waveguide. It is envisioned that, by using structural optimization techniques, the spatial tailoring of the bandgap could be implemented over the 360° span of the lens to achieve full (and almost continuous) control on the direction of the beam. The directivity plots also provide further insight into the formation of the beam highlighting the complete absence of side lobes, which are instead typical of PA excitation. These plots also show that a proper design can achieve zero cross-talk between channels. A minor crosstalk effect is visible in the actuation of channel #4 (producing crosstalk with channel #3).

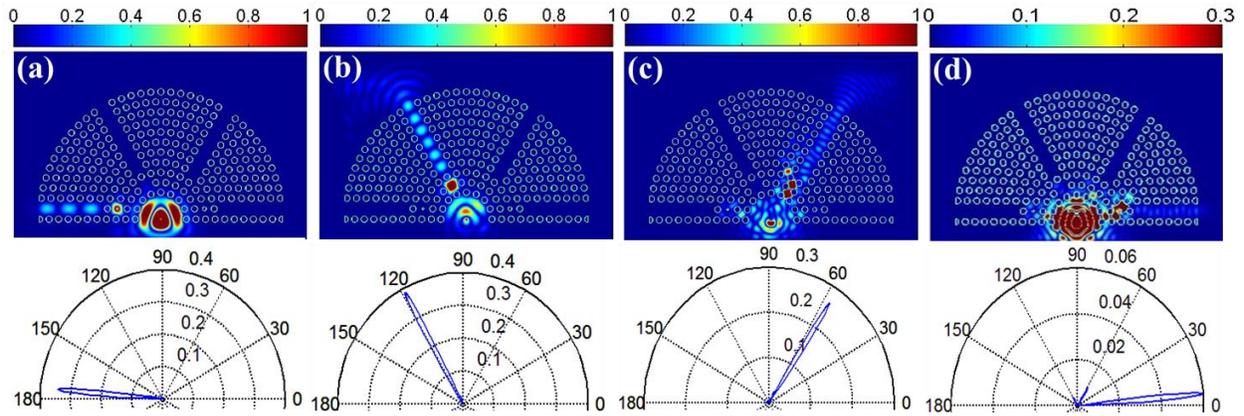

FIG. 4. Numerical simulations showing the dynamic response of the lens when actuated at different driving frequencies. The maps show (top) the squared amplitude of the out-of-plane particle displacement normalized with respect to the driving point displacement $\left|\frac{u_3}{\max(u_{3\_exc})}\right|^2$ and (bottom) the corresponding directivity plots. Results indicate that the non-resonant lens design is able to achieve beam-forming and beam-steering by exploiting a single transducer. (a)-(d) show that by tuning the frequency of the excitation the ultrasonic beam can be steered in selected directions. The four channels are activated at $f_1$=20.87 kHz, $f_2$=24.96 kHz, $f_3$=42.35 kHz, and $f_4$=41.17 kHz.

The drop-channel based lens design was experimentally validated on a 24" × 24" × 0.16" aluminum plate. The lens was realized by means of through the thickness notches (see zoom-in FIG. 5b). This approach greatly simplified the manufacturing process while still creating large impedance mismatch at the interface between the host and the inclusions. The diameter of the notches was chosen to be $r$ =3/64" while the spacing between two consecutive notches in the radial and azimuthal direction $R_r$ and $R_\theta$ were both set at 6.7 mm. The resonant cavities were implemented by a missing notch and three smaller notches (having radius of 1/32", 3/128", and 1/64"), respectively. The aluminum panel (FIG. 5a) was mounted in a metal frame providing fixed-free-fixed-free boundary conditions. Viscoelastic damping tape was also applied to the plate perimeter in order to reduce the effect of reflected waves. The excitation was provided by a piezoelectric shaker attached to the center of the lens and generating out-of-plane displacement. The response of the plate was acquired using a scanning Laser Vibrometer. In order to experimentally indentify the activation frequency of the channels the lens was initially excited using white noise input. The frequency response spectrum was acquired in each channel in order to identify the frequencies with the largest spectral amplitude. Experimental displacement fields (FIG. 5c) clearly show that the different channels can be activated by selecting the proper frequency. Channel #2 and #3 show some crosstalk due to a limited separation between the activation frequencies provided by the defect modes selected for the two channels. This behavior could be limited or completely eliminated by optimizing the design of the drop-channels and of the cavity modes.

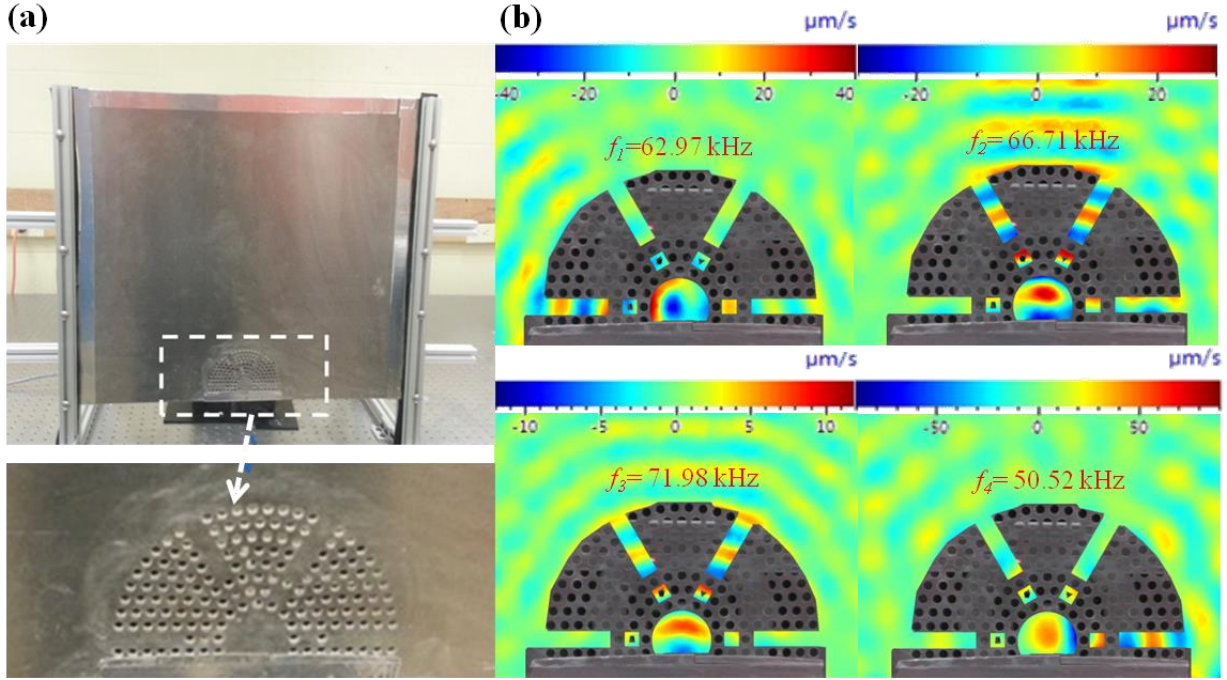

FIG. 5. Experimental setup represented by a thin aluminum panel clamped on two sides. (a) The lens is fabricated by through-holes and it is located near the lower boundary. (b) provides a zoom-in of the lens section. (c) Displacement field maps, obtained via a Scanning Laser Vibrometer, showing the activation of the different channels by tuning the excitation to different frequencies.

Overall, the results show that the lens design is able to create directional and steerable excitation by using a single ultrasonic transducer. The selection of the excitation direction can be simply selected by tuning the frequency of excitation to the activation value for the corresponding defect.

Note that a fundamental difference exists between the performance of the resonant [4] and non-resonant designs. The ultrasonic beam generated by the non-resonant lens design is not collimated and is subjected to diffraction at the exit of the waveguide. In a similar way, the current design cannot provide focusing at a specific structural location. These are limitations that must be considered when selecting the acoustic lens design. Based on the results proposed in this Letter and on the characteristics of resonant lenses illustrated in [4], it is envisioned that the most effective design of the acoustic lens will combine both resonant and non-resonant elements. In particular, the drop-channel design can be exploited to build the overall structure of the lens as well as the mechanism to steer the beam. In addition, a small number of resonant elements can be carefully integrated inside the waveguide in order to provide hyperbolic equi-frequency-contours that are at the basis of ultrasonic focusing and self-collimation [4]. Further studies are obviously needed to explore this hybrid design.

In conclusion, this Letter presented the concept of a structural embedded acoustic lens based on non-resonant metamaterial design. The lens exploits the concept of drop-channel and its design is validated via both numerical and experimental results. The current design offers a much lower fabrication complexity than its resonant counterpart while still achieving directional and steerable interrogation from a single transducer. It is envisioned that both resonant and non-resonant designs could be combined in order to achieve a directional, steerable, self-collimating acoustic lens with largely reduced fabrication complexity.